\documentclass[english]{article}
\usepackage[T1]{fontenc}
\usepackage[latin9]{inputenc}
\usepackage[letterpaper]{geometry}
\usepackage{amsmath}
\usepackage{graphicx}
\usepackage{amssymb}

\makeatletter
\newcommand{\lyxaddress}[1]{
\par {\raggedright #1
\vspace{1.4em}
\noindent\par}
}

\@ifundefined{showcaptionsetup}{}{%
 \PassOptionsToPackage{caption=false}{subfig}}
\usepackage{subfig}
\makeatother

\usepackage{babel}

\begin{document}

\title{Spinless fermion model on diamond chain}

\author{Onofre Rojas\footnote{email: ors@dex.ufla.br; phone: +5535 38291954; fax: +5535 38291961.}
and S. M. de Souza}

\maketitle

\lyxaddress{Departamento de Ciencias Exatas, Universidade Federal de Lavras,
CP 3037, 37200000, MG, Brazil.}
\begin{abstract}
The decoration or iteration transformation was widely applied to solve
exactly the magnetic spin models in one-dimensional and two-dimensional
lattice. The motif of this letter is to extend the decoration transformation
approach for models that describe interacting electron systems instead
of spin magnetic systems, one illustrative model to be studied, will
be the spinless fermion model on diamond chain. Using the decoration
transformation, we are able to solve this model exactly. The phase
diagram of this model was explored at zero temperature as well as
the thermodynamics properties of the model for any particle density.
The particular case when particle-hole symmetry is satisfied was also
discussed.
\end{abstract}
\textbf{Keywords}: decoration transformation, spinless fermion, exactly
solvable models.

\section{Introduction}

Exactly solvable models in statistical physics and mathematical physics
is one of the most challenging topics. Recently several exactly solvable
models were studied in quasi-one dimensional classical-quantum models
such as Ising-Heisenberg models\cite{canova2009,jascur-2004,o-vadim,valverde-2008,canova06,stefano-vadim},
as well as two-dimensional Ising models and Ising-Heisenberg models\cite{valverde-2009,strecka-pre-09},
that can be mapped onto exactly solvable vertex models. Although the
decoration transformation was introduced in the fifty decade\cite{syozi,Fisher},
in order to study the decorated spin models. Due to a successful application
of this approach, recently this method was extended for a general
case of decoration transformation\cite{PhyscA-09}, later extended
by Strecka\cite{strecka-pla} even for classical-quantum models such
as Ising-Heisenberg models. Recently another interesting application
of decoration transformation was also investigated by Pereira \textit{et
al}.\cite{Per-moura} where they considered a delocalized interstitial
electrons on diamond-like chain and they also investigate the magnetocaloric
effect in kinetically frustrated diamond chain\cite{lyra2009}, meanwhile
Strecka \textit{et al.}\cite{streck-elctron} discussed the localized
Ising spins and itinerant electrons in two-dimensional models, as
well as two-dimensional spin-electron with coulomb repulsion\cite{strecka-elect-2010}.

On the other hand studies of strongly correlated electron systems
are certainly one of the most important areas of condensed matter
physics. A large number of papers have been concerned with the investigation
of heavy-fermion behavior, magnetism of strongly correlated systems,
high temperature superconductivity, or metal\textendash{}insulator
transitions and charge-ordering phenomena. A limiting case of these
systems could be the spinless fermion model, where the spin orientation
could be ignored or it can be understand as a fully polarized system.
The one-dimensional case of this model already was investigated by
Czart\cite{czart} and Zhuravlev \textit{et al.}\cite{zhuravlev}.
Although spinless fermion models can be transformed onto XXZ models
in a magnetic field parallel to the anisotropy\cite{czart,zhuravlev}
by the use of the well known Jordan-Wigner transformation, this kind
of mapping could be more involving for the Hubbard-like systems, where
the spin orientation is considered. Therefore, the purpose here is
how to use the decoration transformation approach for interacting
electron systems, without mapping onto spin models. 

The outline of this letter is as follow: In sec. 2 we introduce the
model to be studied. In sec. 3 we present the phase diagram at zero
temperature. In sec. 4 is devoted the decoration transformation and
transfer matrix approach, in order to obtain its exact solution. Whereas
in sec. 5 is discussed the thermodynamics, particle density and correlation
function of the model considered, and finally in sec. 6 we present
our conclusions.

\section{The model}

\begin{figure}
\includegraphics[scale=0.4]{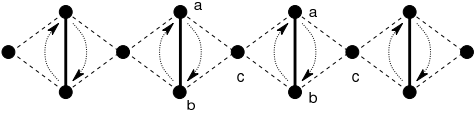}

\caption{\label{fig:Schm-chain}Schematic representation for spinless fermion
model on diamond chain, by vertical solid line we represent a hopping
term and Coulomb repulsion between two sites, while by dashed line
we represent only the Coulomb repulsion term.}

\end{figure}

The model we consider in this letter is the spinless fermion model
on diamond chain as displayed in fig. \ref{fig:Schm-chain}, where
by vertical solid line we mean the particle hopping term and coulomb
repulsion term, while by dashed line we represent only the coulomb
term repulsion. Therefore the Hamiltonian for spinless fermion model
on diamond chain (fig. \ref{fig:Schm-chain}), can be expressed by

\begin{equation}
\mathcal{H}=\sum_{i=1}^{N}\boldsymbol{H}_{i,i+1},\label{eq:Hamt-1}\end{equation}
with $N$ being the number of unit cell, whereas $\boldsymbol{H}_{i,i+1}$
is given by

\begin{align}
\boldsymbol{H}_{i,i+1}= & -t\left(\boldsymbol{a}_{a,i}^{\dagger}\boldsymbol{a}_{b,i}+\boldsymbol{a}_{b,i}^{\dagger}\boldsymbol{a}_{a,i}\right)-\mu\left(\boldsymbol{n}_{a,i}+\boldsymbol{n}_{b,i}+\frac{1}{2}(\boldsymbol{n}_{c,i}+\boldsymbol{n}_{c,i+1})\right)\nonumber \\
 & +V_{1}\boldsymbol{n}_{a,i}\boldsymbol{n}_{b,i}+\frac{V}{2}\left(\boldsymbol{n}_{c,i}+\boldsymbol{n}_{c,i+1}\right)\left(\boldsymbol{n}_{a,i}+\boldsymbol{n}_{b,i}\right),\label{eq:Ham-dmnd-ch}\end{align}
where $\boldsymbol{a}_{\alpha,i}(\boldsymbol{a}_{\alpha,i}^{\dagger})$
are Fermi annihilation (creation) operators for spinless (or completely
polarized) fermion respectively, with $\alpha=\{a,b,c\}$, while $\boldsymbol{n}_{\alpha,i}=\boldsymbol{a}_{\alpha,i}^{\dagger}\boldsymbol{a}_{\alpha,i}$
being number operators. The Hamiltonian parameter $t$ is the hopping
term (solid lines), $\mu$ is the chemical potential, and the Coulomb
repulsion term between fermions on neighboring sites (dashed lines)
is represented $V$ and $V_{1}$. We also consider a periodic boundary
condition.

The spinless fermion model on diamond chain, has an important particle\textendash{}hole
symmetry\cite{Sznajd}. For the purpose of this discussion, the following
canonical transformation $\boldsymbol{a}_{\alpha,i}^{\dagger}\rightarrow\boldsymbol{a}_{\alpha,i}$
and $\boldsymbol{a}_{\alpha,i}\rightarrow\boldsymbol{a}_{\alpha,i}^{\dagger}$
was considered. For the occupation number operator the transformation
reads $\boldsymbol{n}_{\alpha,i}\rightarrow\boldsymbol{a}_{\alpha,i}\boldsymbol{a}_{\alpha,i}^{\dagger}=1-\boldsymbol{n}_{\alpha,i}$,
thus the transformed Hamiltonian becomes\begin{align}
\boldsymbol{H'}_{i,i+1}= & t\left(\boldsymbol{a}_{a,i}^{\dagger}\boldsymbol{a}_{b,i}+\boldsymbol{a}_{b,i}^{\dagger}\boldsymbol{a}_{a,i}\right)-(V+V_{1}-\mu)\left(\boldsymbol{n}_{a,i}+\boldsymbol{n}_{b,i}\right)-(V-\frac{\mu}{2})(\boldsymbol{n}_{c,i}+\boldsymbol{n}_{c,i+1})\nonumber \\
 & +V_{1}\boldsymbol{n}_{a,i}\boldsymbol{n}_{b,i}+\frac{V}{2}\left(\boldsymbol{n}_{c,i}+\boldsymbol{n}_{c,i+1}\right)\left(\boldsymbol{n}_{a,i}+\boldsymbol{n}_{b,i}\right)+\left(V_{1}+2V-3\mu\right).\label{eq:sym-part-hol}\end{align}

The different signs of the hopping term on sites $a$ and $b$ can
be compensated by a further canonical transformation as follows. By
formally replacing say $\boldsymbol{a}_{a,i}^{\dagger}\rightarrow-\boldsymbol{a}_{a,i}^{\dagger}$
($\boldsymbol{a}_{a,i}\rightarrow-\boldsymbol{a}_{a,i}$) for $a$
sites, but keeping unchanged the operators on sites $b$, or vice-versa.

On the other hand, using the Jordan-Wigner transformation it is possible
to map the spinless fermion diamond chain onto a Ising-Heisenberg
diamond chain with non-uniform external magnetic field, as discussed
in reference\cite{valverde-2008,canova06}, the $z$-component of
the spin-1/2 operator is related to number operator by $n=\sigma^{z}+1/2$,
while the creation (annihilation) operators are related through $a_{j}^{\dagger}=2^{j}\left(\prod_{k<j}\sigma_{k}^{z}\right)\sigma_{j}^{-}$
and $a_{j}=2^{j}\left(\prod_{k<j}\sigma_{k}^{z}\right)\sigma_{j}^{+}$.
Therefore the models are related through the relations $J_{H}=V_{1}$,
$J_{I}=V$, $H_{H}=\mu-(V+V_{1})/2$, $H_{I}=\mu-V$ and $J_{H}\Delta=-t/2$. 

The Hamiltonian of spinless fermion model on diamond chain, has not
been discussed yet anywhere . In this sense discussing this model,
we could be opening several variants of fermion models and the most
interesting models are the extended Hubbard-like models\cite{pincus}
with decorated interactions, where is taking into account the spin
orientation, that is why it is interesting first to discuss the spinless
fermion model.

\section{The phase diagram}

Considering the state vectors, where hopping term $t$ is acting at
sites $a$ and $b$. In each state there are only two possibilities
labeled by 0 or 1, it means an empty or occupied particle state respectively,
this leads to the following state vector $|\Psi\rangle=c_{1}|0,0\rangle+c_{2}|0,1\rangle+c_{3}|1,0\rangle+c_{4}|1,1\rangle$,
where $c_{i}$ are the coefficients to be determined for each state.

The states acting on Hamiltonian \eqref{eq:Ham-dmnd-ch} at sites
$a$ and $b$ for each elementary cell becomes, 

\begin{align}
\boldsymbol{H}_{i,i+1}|0,0\rangle= & -\frac{\mu}{2}\left(\boldsymbol{n}_{c,i}+\boldsymbol{n}_{c,i+1}\right)|0,0\rangle,\label{eq:00}\\
\boldsymbol{H}_{i,i+1}|0,1\rangle= & \left[\left(\frac{V}{2}-\frac{\mu}{2}\right)\left(\boldsymbol{n}_{c,i}+\boldsymbol{n}_{c,i+1}\right)-\mu\right]|0,1\rangle-t|1,0\rangle,\label{eq:01}\\
\boldsymbol{H}_{i,i+1}|1,0\rangle= & \left[\left(\frac{V}{2}-\frac{\mu}{2}\right)\left(\boldsymbol{n}_{c,i}+\boldsymbol{n}_{c,i+1}\right)-\mu\right]|1,0\rangle-t|0,1\rangle,\label{eq:10}\\
\boldsymbol{H}_{i,i+1}|1,1\rangle= & \left[\left(V-\frac{\mu}{2}\right)\left(\boldsymbol{n}_{c,i}+\boldsymbol{n}_{c,i+1}\right)+V_{1}-2\mu\right]|1,1\rangle.\label{eq:11}\end{align}

After diagonalize the above $4\times4$ matrix we have 4 eigenvalues,
these eigenvalues depends only on the number of operators $\boldsymbol{n}_{c,j}$,
along with the Hamiltonian parameters. The states \eqref{eq:00} and
\eqref{eq:11} are already in their eigenstates, while the eigenvalues
of eqs. \eqref{eq:01} and \eqref{eq:10} is given simply by $\left[\left(\frac{V}{2}-\frac{\mu}{2}\right)\left(\boldsymbol{n}_{c,i}+\boldsymbol{n}_{c,i+1}\right)-\mu\right]\pm t$,
whereas their respective eigenvectors are given by $|v\rangle_{s,a}=\frac{1}{\sqrt{2}}\left(|0,1\rangle\pm|1,0\rangle\right)$,
symmetric and anti-symmetric states respectively. Although there are
16 eigenvalues only four possible ground states energies were found,
whose eigenvectors for different phases are given by

\begin{align}
|S0\rangle= & \prod_{i=1}^{N}|0,0\rangle_{i}\otimes|0\rangle_{i}, & \rho=0,\\
|S1\rangle= & \prod_{i=1}^{N}\tfrac{1}{\sqrt{2}}\left(|1,0\rangle_{i}+|0,1\rangle_{i}\right)\otimes|0\rangle_{i}, & \rho=1,\\
|S2\rangle= & \prod_{i=1}^{N}\tfrac{1}{\sqrt{2}}\left(|1,0\rangle_{i}+|0,1\rangle_{i}\right)\otimes|1\rangle_{i}, & \rho=2,\\
|S3\rangle= & \prod_{i=1}^{N}|1,1\rangle_{i}\otimes|1\rangle_{i} & \rho=3,\end{align}
 where the states of type $|a,b\rangle\otimes|c\rangle$ corresponds
to the particles states at sites $a$, $b$ and $c$, respectively.$ $

\begin{figure}

\includegraphics[scale=0.3]{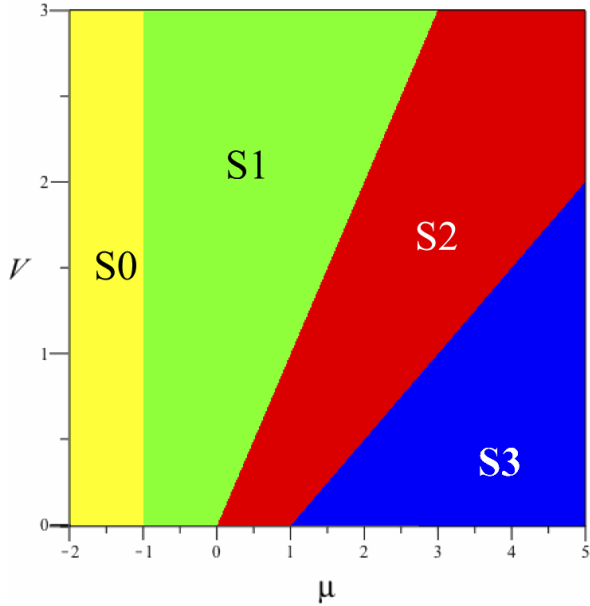}\caption{\label{fig:Ph-dgm}Phase diagram of spinless fermion model on diamond
chain as a function of $\mu$ and $V$, for fixed values of $t=1$
and $V_{1}=V$.}

\end{figure}

The ground state phase diagram of $V$ as a function of $\mu$ is
displayed in fig. \ref{fig:Ph-dgm}, for fixed values of $t=1$ and
$V_{1}=V$. The four states are limited as follow: 

\begin{equation}
|S0\rangle:\;\text{\text{{limited}}}\:\text{{by}}\quad\mu\leqslant-1,\;\text{{with}}\quad\rho=0,\end{equation}

\begin{equation}
|S1\rangle:\;\text{\text{{between}}}\quad\mu\geqslant-1\;\text{{and}}\; V=\mu\quad\text{{with}}\quad\rho=1,\end{equation}

\begin{equation}
|S2\rangle:\;\text{\text{{between}}}\quad V=\mu\;\text{{and}}\; V=(\mu-1)/2\quad\text{{with}}\quad\rho=2,\end{equation}

\begin{equation}
|S3\rangle:\;\text{\text{{between}}}\quad V=(\mu-1)/2\;\text{{and}}\; V=0\quad\text{{with}}\quad\rho=3.\end{equation}

The first one is $|S0\rangle$, this corresponds just to a simple
empty lattice particle (or fully-filled holes) on diamond chain with
total density $\rho=0$ (yellow region). The second state is represented
by $|S1\rangle$, where one particle is at site $a$ or $b$ of the
elementary cell, then the particle density for this state is $\rho=1$
(green region). Furthermore there is another state $|S2\rangle$,
this state corresponds to the configuration that one particle is fluctuating
at sites $a$ and $b$, whereas on site $c$ of elementary cell there
is another particle, then the total density is $\rho=2$ (red region).
Finally there is a fully-filled particle diamond chain state represented
by $|S3\rangle$ with density $\rho=3$, or it can also be understood
as an empty lattice of hole state (blue region).

\section{Decoration transformation and transfer matrix}

In order to study the thermodynamics properties of the spinless fermion
model on diamond chain, it will be used the decoration transformation\cite{Fisher,PhyscA-09,strecka-pla}
as described bellow. Actually, it is not necessary to map the spinless
fermion model onto spin models, like Ising-Heisenberg model\cite{canova2009,jascur-2004,o-vadim,valverde-2008,stefano-vadim}.
Thus, there is an interest in performing the decoration transformation
for operators. In this sense it will be apply directly the decoration
transformation approach\cite{Fisher,PhyscA-09,strecka-pla} for the
proposed model. The main aim to solve the Hamiltonian \eqref{eq:Hamt-1}
of spinless fermion model on diamond chain is to map onto an effective
spinless fermion model without hopping term, whose Hamiltonian is
given by

\begin{equation}
\mathcal{\tilde{H}}=\sum_{i=1}^{N}\left[\tilde{V}\boldsymbol{n}_{c,i}\boldsymbol{n}_{c,i+1}-\frac{\tilde{\mu}}{2}(\boldsymbol{n}_{c,i}+\boldsymbol{n}_{c,i+1})\right],\end{equation}
where $\tilde{V}$ and $\tilde{\mu}$ are coefficients to be determined
using decoration transformation\cite{Fisher,PhyscA-09}.

The Boltzmann factor of effective spinless fermion model can be expressed
as follow

\begin{equation}
\tilde{w}(\boldsymbol{n}_{c,i},\boldsymbol{n}_{c,i+1})=\exp\left(\beta\tilde{V}\boldsymbol{n}_{c,i}\boldsymbol{n}_{c,i+1}-\beta\frac{\tilde{\mu}}{2}(\boldsymbol{n}_{c,i}+\boldsymbol{n}_{c,i+1})\right),\end{equation}
where $\beta=1/kT$, with $k$ being the Boltzmann constant and $T$
the absolute temperature.

On the other hand the Boltzmann factors for the spinless fermion model
on diamond chain given for the Hamiltonian \eqref{eq:Hamt-1} reads
as

\begin{equation}
w(\boldsymbol{n}_{c,i},\boldsymbol{n}_{c,i+1})=\mathrm{tr}_{a,b}\left(\mathrm{e}^{-\beta\boldsymbol{H}_{i,i+1}}\right),\end{equation}
the operators $\boldsymbol{n}_{c,i}$ ranges from 0 to 1, then we
have explicitly the Boltzmann factors,

\begin{align}
w(0,0)= & 1+2\mathrm{e}^{\beta\mu}\cosh(\beta t)+\mathrm{e}^{-\beta V_{1}+2\beta\mu},\\
w(0,1)= & \mathrm{e}^{\beta\mu/2}+2\mathrm{e}^{\beta3\mu/2-\beta V/2}\cosh(\beta t)+\mathrm{e}^{-\beta(V+V_{1})+5\beta\mu/2},\\
w(1,1)= & \mathrm{e}^{\beta\mu}+2\mathrm{e}^{\beta2\mu-\beta V}\cosh(\beta t)+\mathrm{e}^{-\beta(2V+V_{1})+3\beta\mu}.\end{align}

For simplicity the Boltzmann factor for diamond chain are denoted
by $w_{0}=w(0,0)$, and $w_{1}=w(1,0)=w(0,1)$ and $w_{2}=w(1,1)$. 

In order to apply decoration transformation we need to impose the
following condition:

\begin{align}
\mathcal{Z} & =f\mathcal{Z}_{eff}.\end{align}

The Boltzmann factor for both systems must be equal in order to satisfy
the decoration transformation\cite{Fisher,PhyscA-09}. Thus there
are three unknown algebraic equations and three unknown parameters,
this algebraic system is solved easily using the decoration transformation
method, whose solutions are written as

\begin{equation}
f=w_{0},\quad\tilde{\mu}=\frac{2}{\beta}\ln\left(\frac{w_{1}}{w_{0}}\right),\quad\tilde{V}=\frac{1}{\beta}\ln\left(\frac{w_{1}^{2}}{w_{2}w_{0}}\right)\end{equation}
where the $f$, $\tilde{\mu}$ and $\tilde{V}$ are expressed as a
functions of the original parameters of the Hamiltonian by means of
$w_{0}$, $w_{1}$ and $w_{2}$.

On the other hand, the Boltzmann factor for effective spinless fermion
model reads as

\begin{equation}
\tilde{w}_{0}=1,\quad\tilde{w}_{1}=\mathrm{e}^{\beta\tilde{\mu}/2}=x,\quad\tilde{w}_{2}=\mathrm{e}^{\beta\tilde{\mu}-\beta\tilde{V}}=x^{2}y,\end{equation}
with being $x=\exp(\beta\tilde{\mu}/2)$ and $y=\exp(-\beta\tilde{V})$. 

In what follows, we are interested to solve the effective spinless
fermion without hoping term (also known as the atomic limit), to solve
this effective model it will be used the transfer matrix method\cite{baxter-bk},
given by

\begin{equation}
T=\left(\begin{array}{cc}
1 & x\\
x & yx^{2}\end{array}\right).\end{equation}
Note that the terms $\tilde{\mu}$ and $\tilde{V}$ can be obtained
in a similar way as were obtained for spin models. The eigenvalues
of transfer matrix are given by $ $$\lambda_{\pm}=\left(1+yx^{2}\pm\sqrt{\left(1-yx^{2}\right)^{2}+4x^{2}}\right)/2$. 

Using the largest eigenvalues $\lambda_{+}$ of the transfer matrix
$T$, we obtain the partition function per site of the model in terms
of the effective spinless fermion model in atomic limit $\mathcal{Z}=f\mathcal{Z}_{eff},$
with $\mathcal{Z}_{eff}=\lambda_{+}$ is the partition function for
effective spinless fermion without hopping term. The partition function
per elementary cell, is expressed in terms of the Boltzmann factors
for the spinless fermion model on diamond chain, may be written

\begin{align}
\mathcal{Z} & =\frac{1}{2}\left(w_{0}+w_{2}+\sqrt{\left(w_{0}-w_{2}\right)^{2}+4w_{1}^{2}}\right).\end{align}

From the partition function of spinless fermion model on diamond chain,
it is possible to obtain the free energy by the relation $\Omega=-kT\ln\mathcal{Z}$.
Once known this result, we are ready to study several physical amounts
such as entropy, specific heat, average energy and so on.

\section{Thermodynamics and correlation functions}

So far, we have not yet specified the particles number of the diamond
chain. In order to obtain the electron density per elementary cell,
we take the derivative of free energy in relation to the chemical
potential of spinless fermion on diamond chain\begin{equation}
\rho=\left.-\frac{\partial\Omega}{\partial\mu}\right|_{\beta}=2\langle\boldsymbol{n}_{a}\rangle+\langle\boldsymbol{n}_{c}\rangle,\label{eq:rho}\end{equation}
here we are assuming the exchange invariance of sites $a$ and $b$.

The particle density of type $c$ can be obtained from effective spinless
fermion model, by the relation\begin{equation}
\langle\boldsymbol{n}_{c}\rangle=\frac{1}{1+\left(\frac{x}{\lambda_{+}-1}\right)^{2}},\label{eq:avg-nc}\end{equation}
whereas the particle density for sites $a$ and $b$ per site, can
be obtained combining eqs. \eqref{eq:rho} and \eqref{eq:avg-nc},
resulting in \begin{equation}
\langle\boldsymbol{n}_{a}\rangle=\frac{\rho-\langle\boldsymbol{n}_{c}\rangle}{2}.\end{equation}
\begin{figure}
\subfloat[$V=1$ and $V_{1}=1$]{\includegraphics[scale=0.22]{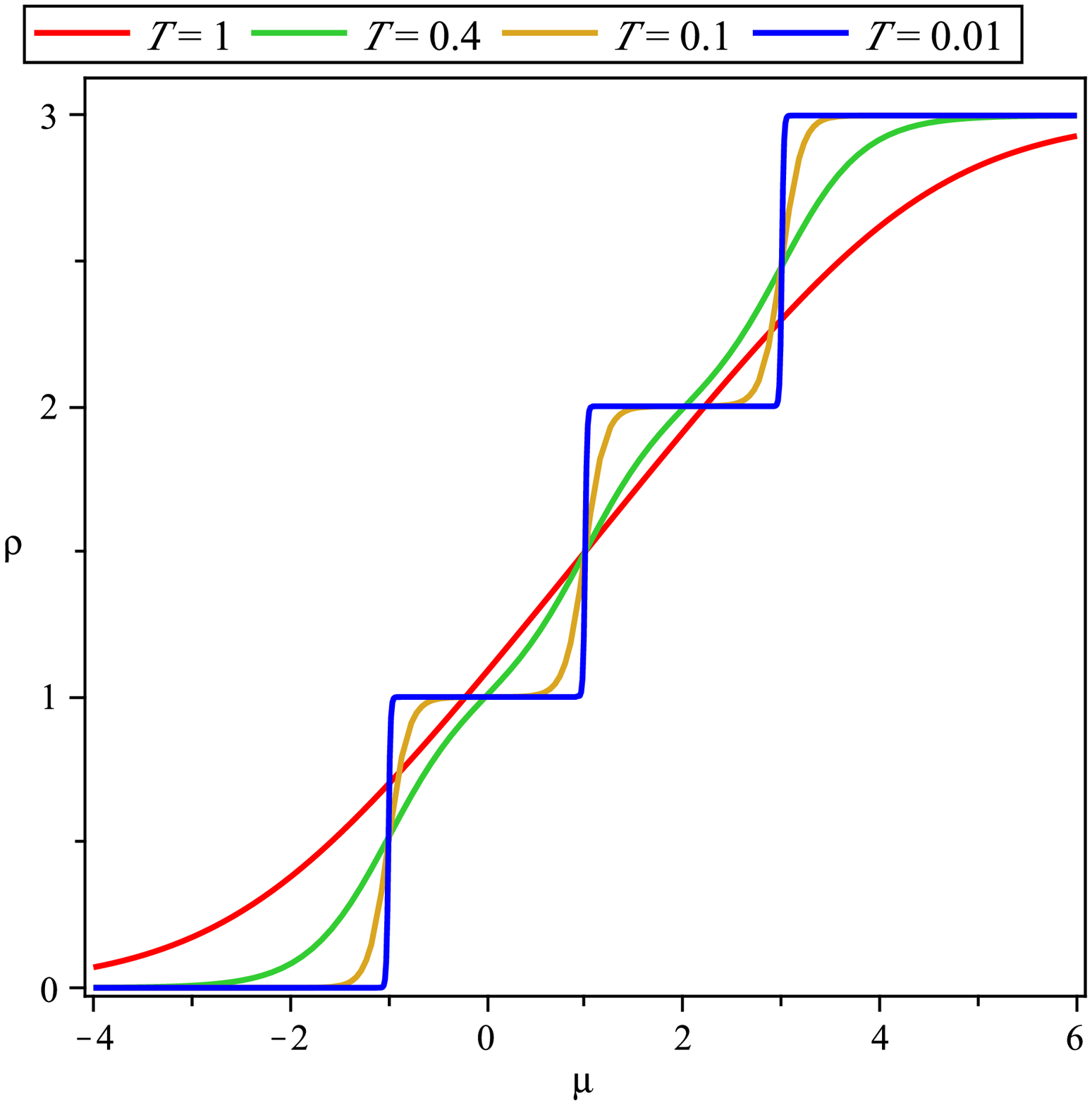}

}\subfloat[$V=1$ and $V_{1}=0.1$]{\includegraphics[scale=0.22]{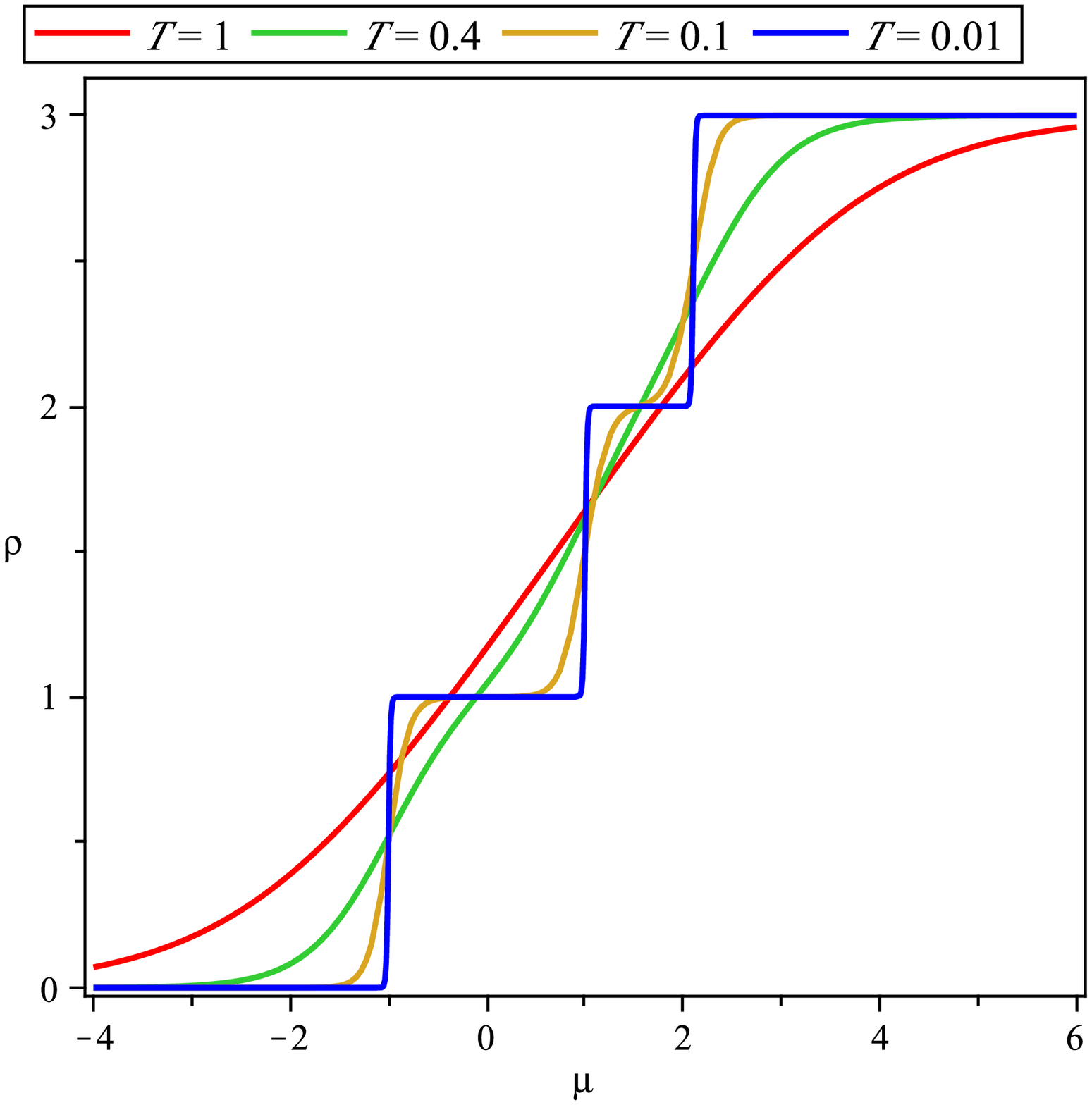}}\subfloat[$ $$V=1$ and $V=1$]{\includegraphics[scale=0.22]{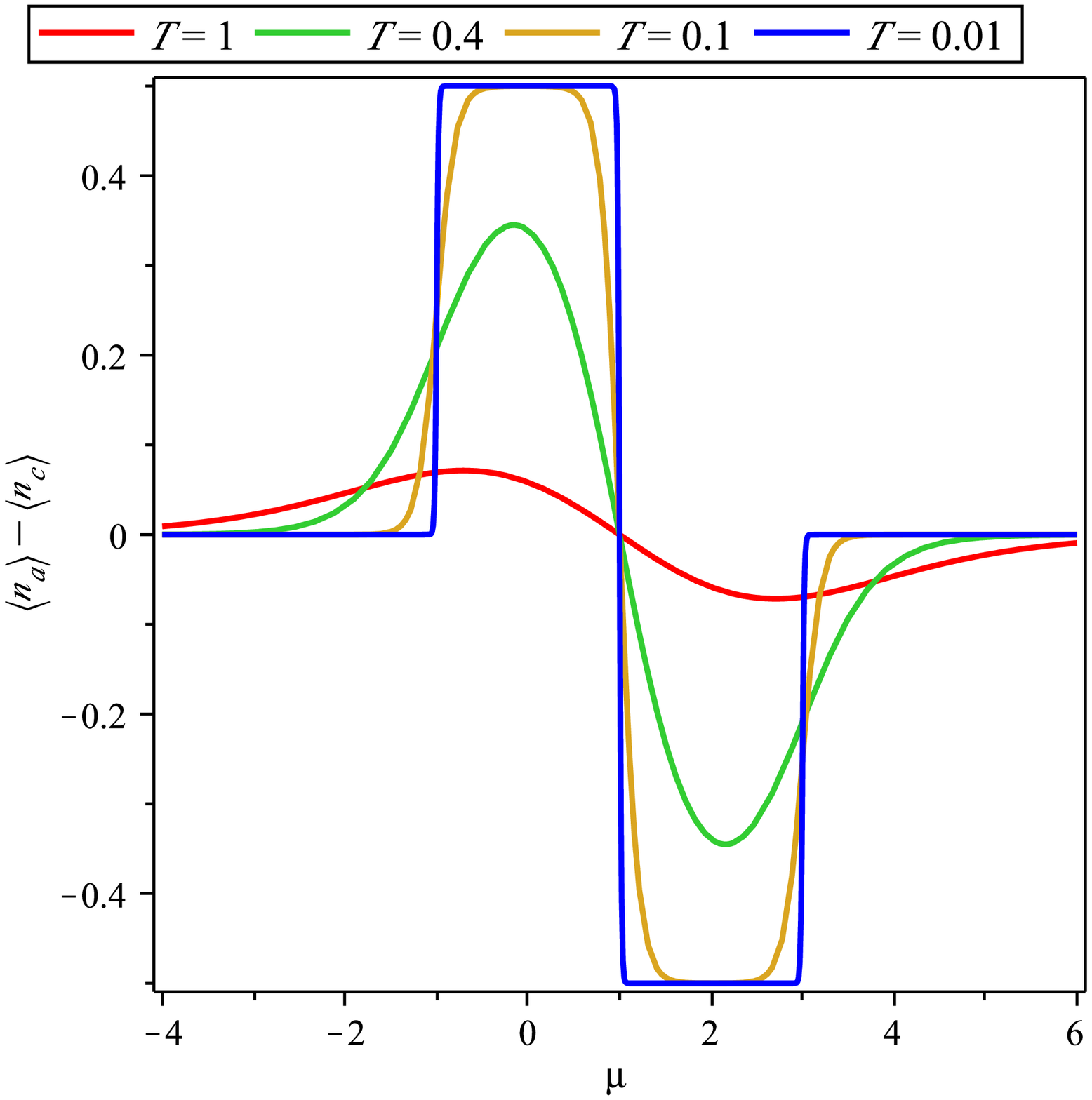}
}\subfloat[$V=1$ and $V_{1}=0.1$]{\includegraphics[scale=0.22]{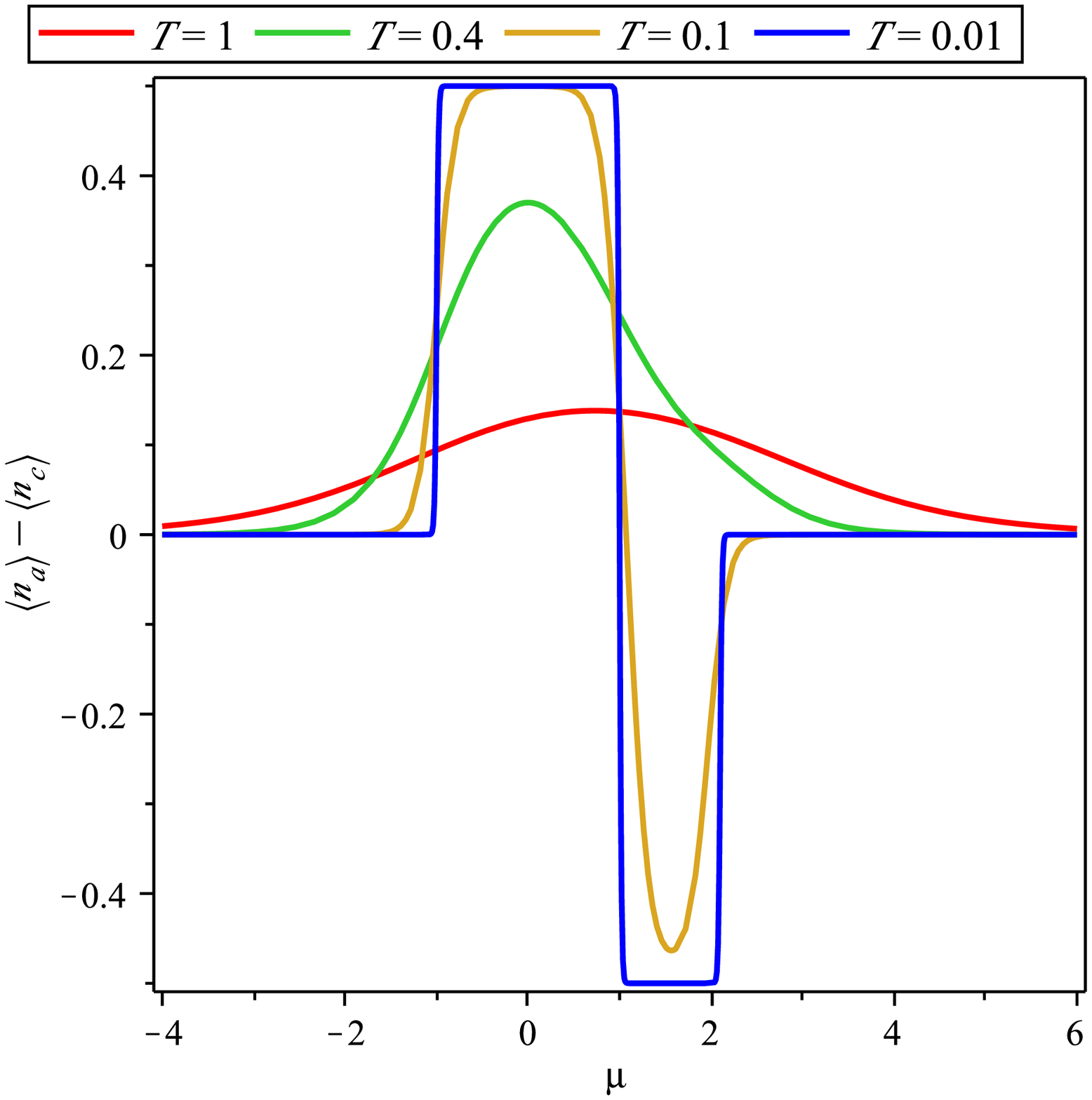}}\caption{\label{fig:dens&mu}In (a) and (b) is plotted the total density $\rho$
as a function of chemical potential $\mu$, for fixed values of temperature.
In (c) and (d) we display the amount $\langle\boldsymbol{n}_{a}\rangle-\langle\boldsymbol{n}_{c}\rangle$
versus $\mu$, for same set of parameters. }

\end{figure}

Further we illustrate the chemical potential behavior for fixed values
of $t=1$, $V=1$ and $V_{1}=\{1,0.1\}$. In fig. \ref{fig:dens&mu}(a-b)
we display the density $\rho$ as a function of chemical potential
$\mu$, for a range of temperatures $T=1.0$, 0.4, 0.1 and 0.01, this
behavior is in agreement with phase transition at zero temperature
(fig. \ref{fig:Ph-dgm}) discussed previously. We show three plateaus
at low temperature, when $\rho$ becomes 1, 2 or 3. The size of plateau
at density $\rho=2$ is proportional the parameter $V_{1}$, for small
$V_{1}$ there is a short plateau (fig.\ref{fig:dens&mu}(b)), while
for large $V_{1}$ long plateau is observed (the last one not illustrated).
In fig. \ref{fig:dens&mu}(c-d) the amount $\langle\boldsymbol{n}_{a}\rangle-\langle\boldsymbol{n}_{c}\rangle$
versus $\mu$ is displayed, in order to study the average particles
number behavior at sites $a$ and $b$ compared to that one on site
$c$. So for negative chemical potential we have $\langle\boldsymbol{n}_{a}\rangle>\langle\boldsymbol{n}_{c}\rangle$,
while for positive chemical potential we have $\langle\boldsymbol{n}_{a}\rangle<\langle\boldsymbol{n}_{c}\rangle$.

Another interesting properties we would like to discuss is the symmetry
particle-hole described by eqs\eqref{eq:sym-part-hol} must be satisfied
only when $V_{1}=V$, therefore we find the following relation for
the free energy,

\begin{equation}
\Omega(t,V,\mu)=\Omega(t,V,2V-\mu)+(3V-3\mu).\end{equation}

In general, the chemical potential depends on the temperature, however
at half-filled band particle density $\rho=1.5$ and under particle-hole
symmetry relation, the chemical potential becomes independent of the
temperature, given simply by $\mu=V$. If we look at the phase diagram
displayed in fig. \ref{fig:Ph-dgm}, the relation $V=\mu$ corresponds
the phase transition between density $\rho=1$ and 2, as expected
the density in this region should be $\rho=1.5$, due to the thermal
fluctuation begins to act.

\begin{figure}

\includegraphics[scale=0.3]{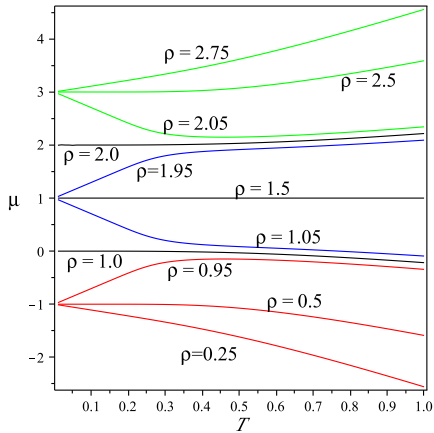}\caption{\label{fig:mu-T}For several values of density $\rho$ is displayed
the chemical potential as a function of temperature, for half-filled
band.}

\end{figure}

Furthermore, in fig.\ref{fig:mu-T} the chemical potential as a function
of the temperature for fixed values of density $\rho$ is displayed,
from this plot we may conclude that there is a chemical potential
independent on the temperature only at half-filled band and when symmetry
particle-hole is satisfied.

Another interesting quantity we would like to comment is the correlation
function, it can also be obtained using transfer matrix method\cite{baxter-bk},
which after some algebraic manipulation we obtain,

\begin{equation}
\langle\boldsymbol{n}_{c,r}\boldsymbol{n}_{c,r+k}\rangle=\langle\boldsymbol{n}_{c}\rangle^{2}+\frac{\langle\boldsymbol{n}_{c}\rangle}{1+\left(\frac{x}{\lambda_{-}-1}\right)^{2}}\left(\frac{\lambda_{-}}{\lambda_{+}}\right)^{k}.\end{equation}

Some other nearest correlation or expected values of higher order
can be obtained, combining the derivative of diamond chain Hamiltonian
with respect to one of its parameter, and the decoration transformation
for correlation function as discussed in reference \cite{Fisher,PhyscA-09}.

\section{Conclusions}

Although the decoration transformation has originally been developed
to study magnetic spin models\cite{Fisher,PhyscA-09,strecka-pla},
and since then, this method was widely applied to solve several decorated
spin models. Motivated by this successful application, this transformation
later was generalized\cite{PhyscA-09}, and recently Strecka\cite{strecka-pla}
extended even for Ising-Heisenberg spin models. However, there is
no any approach developed for electron interacting systems, such as
strongly correlated electron systems. In this sense, we discuss how
the decoration transformation can also be used to map from one decorated
electron system onto another effective electron system. As an illustrative
application of decoration transformation for electron interacting
models, it has been considered the spinless fermion model on diamond
chain, where vertical solid line corresponds to hopping term and repulsive
Coulomb interaction term, while by dashed line we mean the repulsive
Coulomb interaction term between the nearest neighbor (fig. \ref{fig:Schm-chain}).
Furthermore, we discuss some properties of this model, such as the
phase diagram at zero temperature, showing four different states with
a given number of particles, changing from empty lattice (or fully-filled
holes) of particles to fully-filled particles (or empty lattice of
holes) on the diamond chain. The thermodynamics of this model, allows
to display the particle density as a function of chemical potential
per elementary cell as well as the chemical potential as a function
of temperature. Finally an additional quantity also was considered
such as the correlation function.

\section*{Acknowledgments}

This work was partially supported by CNPq and FAPEMIG.

\end{document}